\documentclass[sigconf]{acmart}

\renewcommand\footnotetextcopyrightpermission[1]{} 
\usepackage{amssymb}
\usepackage{graphicx}
\usepackage{wrapfig}
\usepackage{color,soul}
\usepackage{subfig}
\usepackage{pifont}
\usepackage{paralist}
\usepackage{url}
\usepackage{subfig}

\newcommand{\incodiceratio}[0]{\emph{In Codice Ratio}}

\newcommand{\Oversegmentation}[0]{Slice-segmentation}
\newcommand{\oversegmentation}[0]{slice-segmentation}

\newcommand{\Polygonalsegmentation}[0]{Jigsaw-segmentation}
\newcommand{\polygonalsegmentation}[0]{jigsaw-segmentation}

\newcommand{\remark}[1]{{#1}}

\begin{document}

\title{Towards Knowledge Discovery from the Vatican Secret Archives. \incodiceratio{} -- Episode 1: Machine Transcription of the Manuscripts.}

\author{Donatella Firmani}
\affiliation{%
	\institution{Roma Tre University}
	\streetaddress{}
	\city{} 
	\state{} 
	\postcode{}
}
\email{donatella.firmani@uniroma3.it}

\author{Marco Maiorino}
\affiliation{
    \institution{Vatican Secret Archives}
    \streetaddress{}
	\city{} 
	\state{} 
	\postcode{}
}
\email{m.maiorino@asv.va}

\author{Paolo Merialdo}
\affiliation{%
	\institution{Roma Tre University}
	\streetaddress{}
	\city{} 
	\state{} 
	\postcode{}
}
\email{paolo.merialdo@uniroma3.it}

\author{Elena Nieddu}
\affiliation{%
	\institution{Roma Tre University}
	\streetaddress{}
	\city{} 
	\state{} 
	\postcode{}
}
\email{elena.nieddu@uniroma3.it}

\fancyhead{}

\begin{abstract}
\incodiceratio{} is a research project to study tools and techniques for analyzing the contents of historical documents conserved in the Vatican Secret Archives (VSA). In this paper, we present our efforts to develop a system to support the transcription of medieval manuscripts. The goal is to provide paleographers with a tool to reduce their efforts in transcribing large volumes, as those stored in the VSA, producing good transcriptions for significant portions of the manuscripts. We propose an original approach based on character segmentation. Our solution is able to deal with the dirty segmentation that inevitably occurs in handwritten documents. We use a convolutional neural network to recognize characters and language models to compose word transcriptions. 
Our approach requires minimal training efforts, making the transcription process more scalable as the production of training sets requires a few pages and can be easily crowdsourced.
We have conducted experiments on manuscripts from the Vatican Registers, an unreleased corpus containing the correspondence of the popes. With training data produced by 120 high school students, our system has been able to produce good transcriptions that can be used by paleographers as a solid basis to speedup the transcription process at a large scale.
\end{abstract}


\maketitle

\section{Introduction}
\label{sec:intro}

\emph{In Codice Ratio} is a research project that aims at developing novel methods and tools to support content analysis and knowledge discovery from large collections of historical documents. The goal is to provide humanities scholars with novel tools to conduct data-driven studies over large historical sources. Paleographers and philologists could quantitatively analyze trends and evolution of writings and languages across time and countries; historians could examine and discover facts and correlations among information spread in vast corpora of documents.

\begin{figure}
    \centering
    \includegraphics[width=0.4\textwidth]{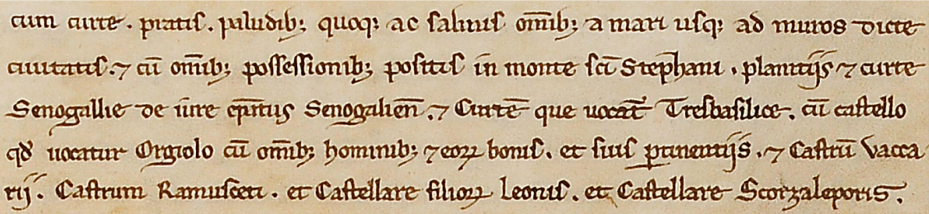}
    \caption{Sample text from the manuscript ``Liber septimus regestorum domini Honorii pope III'' (Vatican Registers).}
    \label{fig:carolina}
\end{figure}

The project concentrates on the collections of the Vatican Secret Archives (VSA), one of the largest and most important historical archive in the world. In an extension of 85 kilometres of shelving, it maintains more than 600 archival collections containing historical documents on the Vatican activities 
--such as, all the acts promulgated by the Vatican, account books, correspondence of the popes--starting from the eighth century. 
We are currently working on the collection of the Vatican Registers, which records the inbound and outbound correspondence of the popes: political letters that testify the broad activities of the popes in the ecclesiastical and temporal spheres; authoritative opinions on legal issues; documents addressed to sovereigns, religious and political institutions scattered throughout the globe; correspondence relating to the harvest of tithes and tributes due to the Church. A small illustration of a page from the Vatican Registers is shown in Figure\ref{fig:carolina}.

The systematic and continuous preservation of these registers began in the middle age,\footnote{Namely, under Pope Innocent III (1198-1216).} hence most of these documents are {\em manuscripts}. 
The VSA has begun to acquire digital images of these documents but, unfortunately, for the earliest registers there not exist complete transcriptions. Therefore, a first fundamental step to develop any form of data-driven content analysis is to perform a transcription of the manuscripts. The problem is challenging: on the one hand, a manual transcription is unfeasible (at least in a reasonable amount of time), due to the volume (hundreds of thousands of pages) of the collection. On the other hand, although these manuscripts are written with a uniform style (a derivation of the \emph{Caroline} style), traditional OCR does not apply here, because of irregularities of writing, ligatures and abbreviations.

\begin{figure}
    \centering
    \includegraphics[height=3cm]{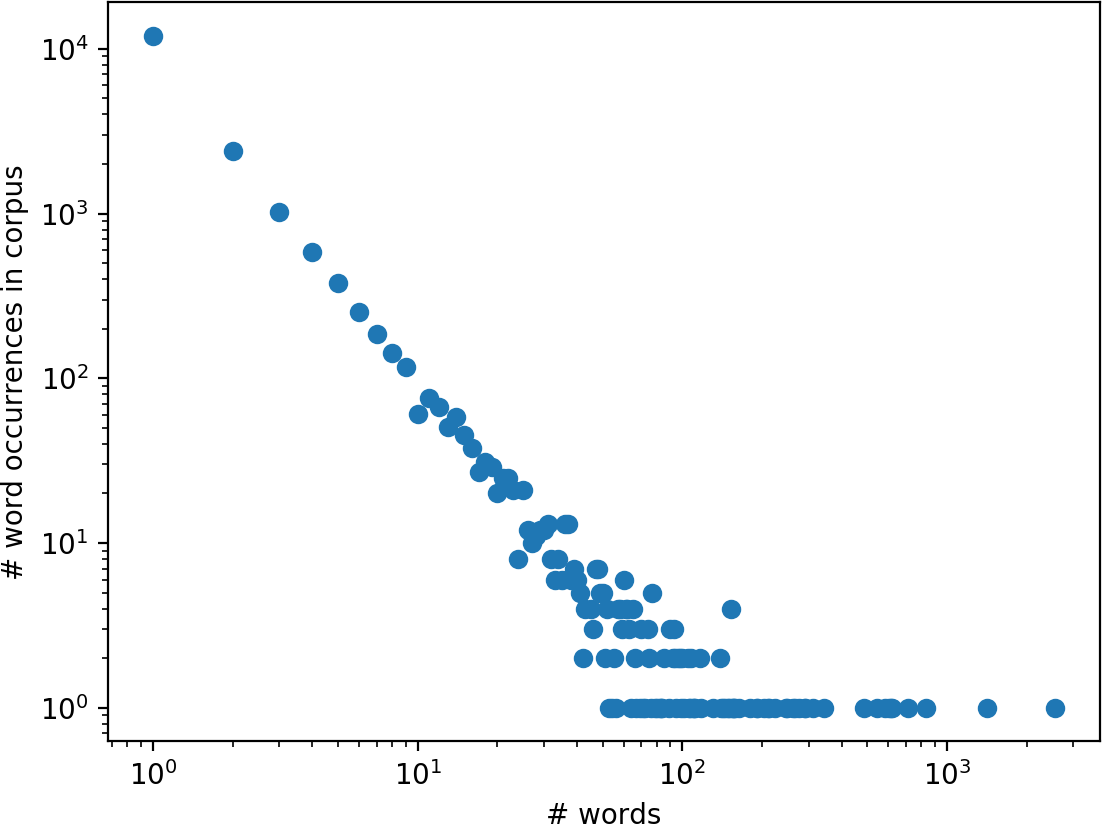}
    \caption{Word distribution in the Registers (log-log scale).} %
    \label{fig:worddist}
\end{figure}

As many libraries and archives have begun digitizing their assets, several handwritten text recognition approaches have been recently developed to address text transcriptions also from medieval manuscripts. Since segmenting words in characters is tricky with hand written texts, these approaches typically aim at recognizing entire words. Because of the variability and the size of the lexicon, they need a huge amount of training data, i.e., hundreds of fully transcribed pages, and often end up focusing on specific tasks, such as {\em word spotting}~\cite{frinken2012novel,almazan2014word}, which aim at recognizing words from a given small vocabulary (e.g. a list of names). To illustrate the problem of building a reliable training dataset, consider Figure~\ref{fig:worddist}: it reports the distribution of the occurrences of words in a corpus composed by a partial transcription of the Registers of Innocent III (in total, it is about 68,000 words). Observe that a few words (just 9) occur more than 100 times (the most occurring word is {\em ''et``}, the Latin conjunction that corresponds to ''and``), while the majority of words have less than 10 occurrences. 

Our goal is to develop a full-fledged system that transcribes as much as possible from the manuscripts, letting the paleographer to complete the produced transcription with local corrections. To this end, we follow a different approach, based on character segmentation. Our idea is to govern imprecise character segmentation by considering that correct segments are those that give rise to a sequence of characters that more likely compose a Latin word. We have therefore designed a principled solution that relies on a convolutional neural network classifier and on statistical language models. For every word, we perform a segmentation that can produce more segments than those actually formed by the characters in the word. Every segment is labeled by a classifier, which essentially recognizes the most likely character. We then organize the sequence of segments in a directed acyclic graph: the paths of such a graph represent candidate transcriptions for the word, and the most likely solution is selected based on language statistics. 

It is worth noting that compared to a segmentation-free approach, training the classifier requires labeled examples for the limited set of character symbols, with a twofold advantage. First, the size of the training set is several order of magnitude smaller, as we need to provide examples only for the limited set of character symbols, and not for a rich lexicon of words. Second, producing the examples is much easier, as it does not require to transcribe whole words, an activity that can be carried on by expert paleographers. In our system, the production of the training set is accomplished by a crowdsourcing solution that consists of simple visual pattern matching tasks, similar to captchas. 

To summarize, our on-going experience for the transcription of the manuscripts of the VSA makes the following contributions:
\begin{itemize}
    \item we have designed an end-to-end solution that  builds on state-of-the-art technologies---such as, convolutional neural network, statistical language models, crowdsourcing---to support paleographers to transcribe large corpora of manuscripts; 
    
    \item we have experimented our approach on the Vatican Registers: never having been transcribed in the past, these documents are of unprecedented historical relevance; we were able to generate the exact transcription for 65\% of the word images of our dataset; for an additional 10\% the correct transcription was at edit distance 2; 

    \item our approach drastically simplifies the production of the training set; in our experiments the training set has been generated in a few hours involving 120 high school students and using just 2 folios of the Vatican Registers;

    \item all the material (code, ground-truth, training data) of the project is publicly online.%
    \footnote{http://www.inf.uniroma3.it/db/icr/}

\end{itemize}

\section{Overview}
\label{sec:overview}

\begin{figure}
    \centering
    \includegraphics[height=1 cm]{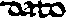}
    \caption{Sample input image of our work-flow. The correct transcription is the Latin word ``dato'' (meaning ``given'').}
    \label{fig:dato}
\end{figure}

Our raw input data consist of high-quality scanned images of whole manuscript \emph{pages}. Each page undergoes a set of standard pre-processing steps, leading to a bag of word images.
\begin{enumerate}
    \item the color image is transformed into a bi-chromatic one;
    \item all the white margins are cropped;
    \item \emph{skew} and \emph{slant} are corrected, that are, page distortions due to acquisition process and letter slope due to calligraphy;
    \item lines are separated out, by looking at horizontal white spaces;
    \item words are extracted, by looking at vertical white spaces. 
\end{enumerate}
Each word image is finally submitted to our transcription system. Figure~\ref{fig:dato} shows a pre-processed word image, of size $66 \times 18$ pixels.

\paragraph{Main components} Our system consists of four main components.
\begin{itemize}
    \item \textbf{Collection of training samples.} We implemented a custom crowd-sourcing platform, and employed 120 high-school students to label the dataset. To overcome the complexity of reading ancient fonts, we provided the students with positive and negative examples of each symbol. We trained a character classifier on this dataset, implemented with a deep convolutional neural network.
    \item \textbf{Character recognition.} Recognizing characters within a handwritten word is challenging, due to \emph{ligatures}. To this end, we first partition the input word into elementary text segments. Most segments contain actual characters, but there are also segments with spurious ink strokes. Then, we submit all the segments to the trained classifier. Computed labels are very accurate when the input segment contains an actual character, but can be wrong otherwise.
    \item \textbf{Candidate transcriptions.} We reassemble noisy labels from the classifier into a set of candidate word transcriptions. Specifically, we select the best $m$ candidate transcriptions for the input word image, using language models.
    \item \textbf{Word decoding.} We consider the $m$ transcriptions from the previous step and revise character recognition decisions in a principled way, by solving a specific decoding problem on a high-order hidden Markov model. The most promising transcriptions are finally returned to the user.
\end{itemize}

\begin{figure}
    \centering
    \includegraphics[height=4cm]{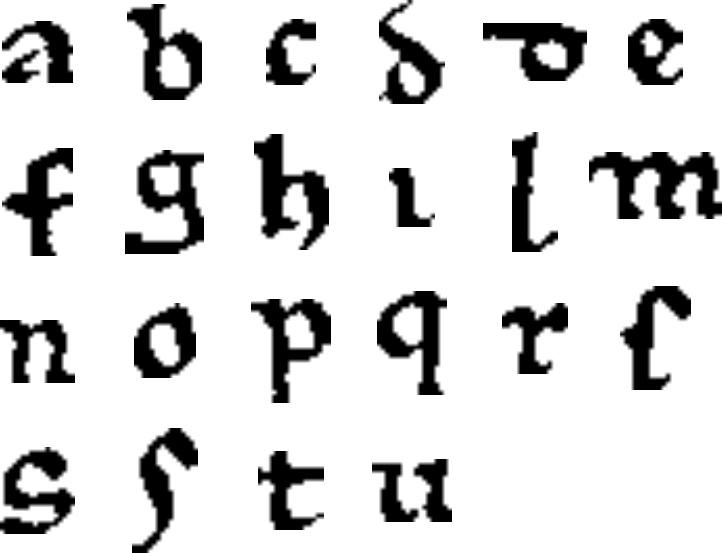}
    \caption{Our set of $22$ symbols. \label{fig:symbols}}
\end{figure}

\paragraph{Character symbols} We take into account minuscule characters of the Latin alphabet and their variations, shown in Figure~\ref{fig:symbols}.

\section{System Work-Flow}
\label{sec:architecture}

In this section, we describe in detail each component of our system. We use the word image in Figure~\ref{fig:dato} as running example.

\subsection{Training samples}
\label{sec:training}

\begin{figure}
    \centering
    \includegraphics[height=3.5cm]{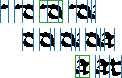}
    \caption{Breakdown of the word in Figure~\ref{fig:dato} into overlapping segments\protect\footnotemark. Local minima are shown in blue. Pieces fitting entirely one character are highlighted in green. \label{fig:boxes}}
\end{figure}

\footnotetext{We show only $10$ out of the $30$ segments from \oversegmentation{}, for lack of space.}

The fundamental step of our transcription system consists in collecting labelled samples of character symbols in Figure~\ref{fig:symbols}. In the HTR domain, there is a common understanding that there is no advanced strategy for partitioning a word image into its character constituents, without being aware of its transcription somehow. Therefore, we do something basic at this point, which turns out to be very effective for our purposes. 
As shown in Figure~\ref{fig:boxes}, we break down a word into overlapping segments of different sizes, bounded by \emph{local minima} of the black pixel count function over columns of the image matrix. Then, we ask the crowd to select only segments in which a given visual pattern (among those in Figure~\ref{fig:symbols}) fits \emph{entirely}. We observed that almost all the ``non-characters'' (including small, large, and misplaced segments) are filtered out, leaving nicely cropped segments such as those highlighted in green in Figure~\ref{fig:boxes}.

We first describe our crowd-sourcing application and then detail our basic segmentation technique, which we call \emph{\oversegmentation{}}.

\begin{figure}
    \centering
    \includegraphics[height=4cm]{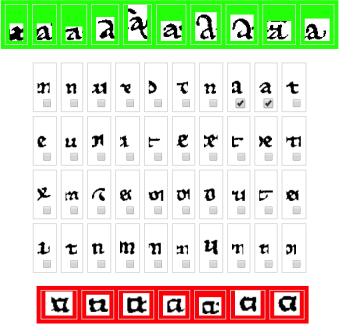}
    \caption{Sample screen of our platform. Positive and negative examples are highlighted in green and red respectively.\label{fig:webapp}}
\end{figure}

\begin{figure}
    \centering
    \includegraphics[height=1 cm]{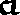}
    \caption{Sample pattern resembling modern hand-writing style of character ``a'', produced by wrong segmentation of the bi-gram ``ci''. Our target pattern of character ``a'' is given by the top-leftmost symbol in Figure~\ref{fig:symbols}.}
    \label{fig:falsefriends}
\end{figure}

\paragraph{Crowd-sourcing} We developed a custom crowd-sourcing platform and enrolled 120 high-school students in the city of Rome, that did the labeling as a part of a work-related learning program. Each student was required to perform a sequence of tasks: each task consists of selecting any image on the platform that visually matched (fitting entirely with least possible extra-strokes) a given character symbol. In Figure~\ref{fig:webapp}, we show a screen-shot of a task. Each task relates to a symbol, and consists specifically of:
\begin{itemize}
    \item sample occurrences of the symbol (in green), best capturing hand-writing variations identified by paleographers;
    \item $40$ images randomly drawn among a pool of unlabelled segments, arranged in a grid, each with its own check-box.
\end{itemize} 
Every time the check-box is marked, the corresponding image receives a vote. Finally, the image is labelled with the most voted character symbol. %
If no clear majority emerges, then the image is labelled with a special \emph{non-character} class $\otimes$. Students were told to match visual patterns, rather than trying to \emph{read}. However, enrolled students do use modern variants of certain symbols in their daily lives, and is reasonable to assume that some tasks have been solved by using reading skills. In order to prevent mislabeling of patterns that casually resemble modern writing styles of symbols, such as Figure~\ref{fig:falsefriends}, we also provided examples of \emph{negative} cases (in red).

\paragraph{\Oversegmentation{}} 
Given the input image word, we count the ink pixels for each column of the matrix encoding the image. Local minima in the resulting function are considered as possible character boundaries. Our intuition is that, typically, ligatures correspond to vertical ``bands'' with less ink. In the case of adjacent local minima -- which can occur if the black pixel count function has plateaus -- only the leftmost candidate point is chosen.

\subsection{Character recognition}
\label{sec:recognition}

\begin{figure}
    \centering
    \includegraphics[height=1 cm]{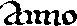}
    \caption{A sample occurrence of the Latin word ``anno'' (meaning ``year'') yielding many OCR interpretations. Interpretations include ``aiiiio'', ``aimo'', ``amio'', ``aniio', ``aiino'', and ``ainio'', but none of these is a Latin word. \label{fig:anno}}
\end{figure}

The core of our transcription system is a character classifier that takes as input a segment of a word image. The classifier estimates the probability that the image belongs to each of the character classes in Figure~\ref{fig:symbols} and to the special non-character class $\otimes$. The classifier is implemented with a \emph{deep convolutional neural network}, and trained with an augmented set of the crowd-sourced image labels. The resulting model, which has been described in~\cite{ai4ch}, achieves $96\%$ accuracy, which is one of the highest results reported in the literature so far\footnote{Average precision and recall, among all classes, are both $96\%$. Precision ranges from $86\%$ to $99\%$, whereas recall ranges from $74\%$ to $99\%$.}. One may think that such an accurate classifier allows the user to just feed it with images from \oversegmentation{} of the input word $w$ and let it filter out almost everything but the nicely cropped segments, which together constitute $w$. However, there are two problems that a transcription system needs to solve.
\begin{itemize}
    \item While humans can easily distinguish character symbols from spurious strokes, this turns out to be a hard task for an automatic classifier. Respective to the class $\otimes$ (non-character), indeed, our classifier achieves $95\%$ precision but only $74\%$ recall. This means that our transcription system needs to deal with a certain amount of ``false characters''.%
    \item Depending on the writing-style, the visual features of certain handwritten characters components, such as the individual sticks of a ``m'', are indistinguishable from different whole characters, such as the ``i'' . In these cases, the ``i''-shaped elements and the whole ``m'' ought to be considered equally valid OCR instances. Only when we consider the word holistically, indeed, we can rule out infrequent cases such as ``iii''. Figure~\ref{fig:anno} provides an extreme example of this phenomenon.
\end{itemize}

\begin{figure}
    \centering
    \subfloat[\oversegmentation{}]{\includegraphics[height=1 cm]{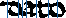}}
    \hspace{0.5cm}
    \subfloat[\label{fig:poly}\polygonalsegmentation{}]{\includegraphics[height=1 cm]{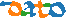}}
    \caption{\Oversegmentation{} vs. \polygonalsegmentation{}.\label{fig:polygonal}}
\end{figure}

Our solution is two-fold.
\begin{itemize}
    \item We introduce a more sophisticated segmentation technique, which we refer to as \emph{\polygonalsegmentation{}}, with several benefits, including reducing ``false characters'' in the recognition phase. The difference between the \oversegmentation{} and the \polygonalsegmentation{} is schematically shown in Figure~\ref{fig:polygonal}. 
    \item We handle ambiguity intrinsic in the character classification process by using a custom version of the so-called \emph{transcription lattice} data-structure, that has been introduced in~\cite{45422} for \emph{on-line} HTR systems. In the on-line setting the handwriting is given as a temporal sequence of coordinates that represents the pen tip trajectory, whereas in our case -- which is typically referred to as \emph{off-line} HTR -- the handwriting is given as an image of the text, without any time sequence information.
\end{itemize}

The \polygonalsegmentation{} is discussed first, then we give details of character classifier. Finally, we describe our lattice data-structure.

\paragraph{\Polygonalsegmentation{}} For each connected component of the input word image, we compute its upper and lower contour\footnote{We apply a smoothing of $3$ pixels to both the resulting functions}. We then compute local minima for the upper contour, and local maxima for the lower contour. Having both functions aligned to the column index of the word image, we connect each local minimum from the upper contour to the closest local maximum from the lower contour. Resulting ``regions'', or segments, can be combined together into larger ones by using the lattice. Figure~\ref{fig:poly} shows regions extracted by the \polygonalsegmentation{} on a sample word, with different colors.

\paragraph{Character classifier} Our classifier is implemented with the deep convolutional neural network described in~\cite{ai4ch}. For sake of completeness, we mention that it takes as input $56 \times 56$ single-channel images. The input is then propagated through $8$ adaptable layers, whose design is inspired to recent neural networks models~\cite{goodfellow2016deep}.

\begin{figure}
    \centering
    \subfloat[\label{grapha} $\eta=1.0$]{\includegraphics[width=0.4\textwidth]{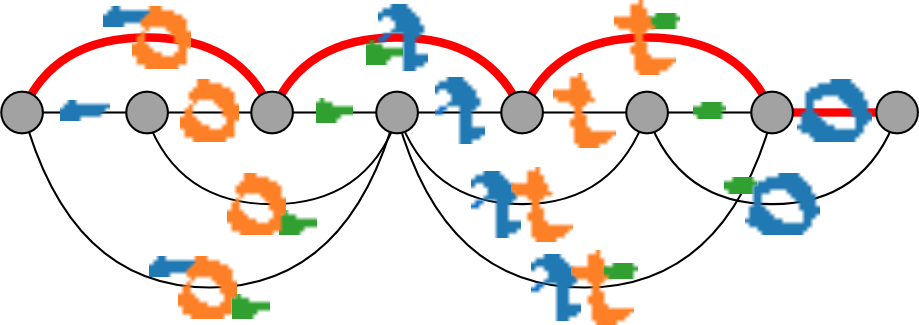}}
    
    \subfloat[$\eta < 1.0$\label{graphb}]{\includegraphics[width=0.4\textwidth]{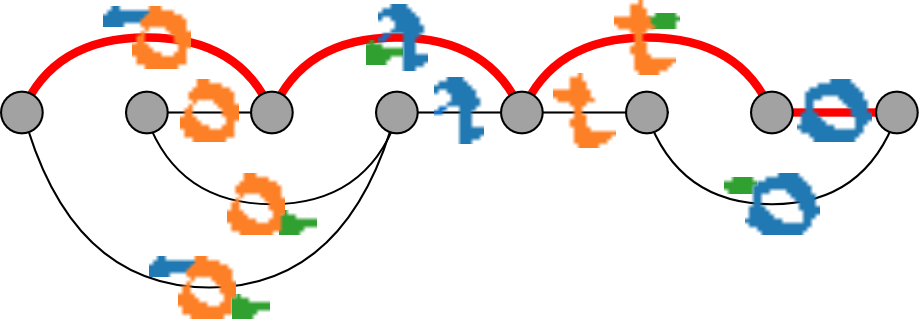}}    
    \caption{Sample lattices with different thresholds.} %
\end{figure}

\paragraph{Transcription lattice} We consider the \polygonalsegmentation{} of the input word, the $x$ coordinates of segment centroids, and a ``start'' centroid corresponding to the $x=0$ coordinate. Using these data, we build a directed acyclic graph where:
\begin{itemize}
    \item each vertex is representative of a centroid $c$;
    \item each edge $(c_1,c_2)$ represents the \emph{group} of adjacent segments whose centroids $c$ are in the interval $(c_1,c_2]$, and possibly form together a character shape.
    \item each path yields a \emph{partition} of the word into groups of adjacent segments (i.e., candidate characters).
\end{itemize}
We use the term ``edge'' for referring both to the mathematical object and to the group of adjacent segments. Edges that span a character-sized length -- that is, that have endpoint centroids not further\footnote{Centroid distance is just $c_2-c_1$.} than a threshold $\sigma$ -- are submitted to our classifier. Then, only those that get a probability score for the class $\otimes$ lower than a threshold $\eta$ are effectively added to the lattice. We set the default values for $\sigma$ and $\eta$ to $25$ pixels (we observed experimentally that edges larger than $25$ pixels can be safely dropped) and $0.1$, respectively.

\begin{example}
Figure~\ref{grapha} shows the complete transcription lattice for our word in Figure~\ref{fig:dato}, with the path yielding the correct partition in red. Seven distinct segments produce eight vertices, one for each segment plus the ``start'' vertex. Edges between vertices represent alternative groupings for the distinct segments. Figure ~\ref{graphb} shows the same graph, without the edges which would likely be discarded by the classifier as $\otimes$. Discarding edges reduces the number of paths, and therefore the number of unlikely partitions for a word.
\label{ex:graph}
\end{example}

\noindent Each edge in the lattice gets an array of \emph{labels}, corresponding to the top-ranked classes from the classifier. Even though the top class is correct in the $96\%$ of the cases, considering a few alternatives from the probability distribution returned by the classifier can be useful in the most difficult instances. Therefore, we allow each edge in the lattice to have multiple labels, using the following heuristic.
\begin{enumerate}
    \item We sort classes according to probability.
    \item We pick the top classes that account together for at least $\theta_1$ probability (that is, the sum of their scores is $\geq \theta_1$), and that account singularly for at least $\theta_2$ (that is, their score is $\geq \theta_2$).
    \item If we select the class $\otimes$ in this process, we drop the edge.
    \item Otherwise, we set selected classes as its labels.
\end{enumerate}
Default values for $\theta_1$ and $\theta_2$ are $0.8$ and $0.1$, respectively \footnote{In practice, we never select more than three labels.}.

\begin{example}
Let $\{l_1: p_1, l_2: p_2, l_3 : p_3\}$ be the top-$3$ classes from the classifier. We show some computations of our heuristic in the default setting for deciding labels in a few sample cases. 
\begin{itemize}
    \item $\{\text{``a''}: 0.8, \text{``o''}: 0.1, \text{``d''}: 0.05\}$ yields a single label ``a''.
    \item $\{\text{``a''}: 0.75, \text{``o''}: 0.05, \text{``d''}: 0.05\}$ also yields a single label ``a'', because for scoring total $0.8$ we would need the low-probability label ``o'' (which violates $\theta_2$).
    \item $\{\text{``a''}: 0.5, \text{``o''}: 0.4, \text{``d''}: 0.05\}$ yields a double label ``a/o''.
\end{itemize}
\label{ex:thresholds}
\end{example}

\subsection{Candidate transcriptions}
\label{sec:generation}

Let the \emph{origin} vertex be the start centroid, and the \emph{destination} vertex be any of the sink vertices (i.e., those with no outgoing edges). Note that one sink vertex corresponds to the rightmost centroid, while possible other sinks are produced by the edge selection process. We generate all the edge-disjoint paths from the origin to all possible destinations: each corresponds to a candidate transcription. An edge with multiple labels can be thought of as a set of \emph{parallel} edges (i.e., a set of edges that have the same endpoints), each with its own label. We generate transcriptions in a depth-search-first fashion, and grow each transcription incrementally from left to right.

\begin{example}
Consider the lattice in Figure~\ref{graphb}. The candidate transcriptions are ``dato'', ``daid'', ``diid'', ``dito''.

\label{ex:paths}
\end{example}

Generated transcriptions are ranked in non-increasing order of \emph{word probability}, based on a \emph{model} of the Latin language. We first describe this step, and then make time-efficiency considerations.

\paragraph{Ranking transcriptions} We use a Latin corpus for estimating $q$-gram frequencies in the Latin language. Probability of each transcription is estimated under the assumption that the frequency of each character only depends from the preceding $q-1$ characters. Consider the transcription ``dato'' in Example~\ref{ex:paths}, and let $q=3$. There are two classical methods for evaluating its probability.
\begin{itemize}
    \item We can think of the transcription as a \emph{sub-string} of a possibly larger word, and compute its \emph{sub-string probability} as $p(dato)=p(d)p(a|d)p(t|da)p(o|at)$.
    \item We can think of the transcription as a \emph{word} of the language,  and compute its \emph{word probability} as $p(\$dato\wedge)=p(d|\$)p(a|\$d)p(t|da)p(o|at)p(\wedge|to)$, where $\$$ and $\wedge$ are special symbols denoting the beginning and the end of a word.
\end{itemize}

\noindent{} Terms in the above formulae are part of the language model. Word probability better represents whether a given transcription corresponds to a Latin word, whereas sub-string probability is more robust to word image extraction failures, where the word image lacks the first (or the last) few symbols. We rank transcriptions according to their word probability, with $q=6$ by default, and use sub-string probability for optimization (as discussed in the following). Some sinks can be close to the start vertex due to noise in the \polygonalsegmentation{}, yielding short transcriptions even for long word images. Such shorter paths can have higher word probability than longer ones, therefore, we introduce a simple yet robust method for getting rid of them.
\begin{enumerate}
    \item We compute word image length.
    \item We estimate the size that each transcription would have on paper, by multiplying the average character length observed in our experiments (that is, $19$ pixels) by the number of characters in the transcription.
    \item We drop all the generated transcriptions with length smaller than $90\%$ of the word image length.
\end{enumerate}

\paragraph{Speed-up} Generating all the paths can be time-consuming, and avoiding uninteresting transcriptions can save time without affecting result quality. For this purpose, we use sub-string probability. Since sub-string probability decreases monotonically as the transcription grows longer, as soon as the sub-string probability of a path from the origin vertex drops under a threshold $\beta$, we ``prune'' all the paths with that prefix from our computation. Even though word probability may have different value than sub-string probability, it is reasonable to assume that low sub-string probability transcriptions correspond to low word probability transcriptions. We set the default value for $\beta$ to $10^{-16}$ (note that path probability is the product of possibly many terms due to chain rule). 

\subsection{Word decoding}

The final step of our work-flow consists in revising character recognition decisions, and possibly making punctual changes to candidate transcriptions. In the case of wrong classifier decisions, we may not have generated the correct transcription of a word image. So-called \emph{decoding} algorithms, such as the Viterbi algorithm~\cite{viterbi}, are one classical approach for solving those cases in HTR and speech-to-text applications. The decoding problem consists indeed in finding the most likely sequence of hidden states of a Hidden Markov Model (HMM), that resulted in a given sequence of observed events. 

In the HTR domain:
\begin{itemize}
    \item the \emph{hidden sequence} is the underlying word in the image;
    \item the \emph{observed sequence} is the the generated transcription from the lattice, with labels computed by the classifier;
    \item the \emph{transition probabilities} are the language model: for instance, frequency of bi-gram ``ci'' corresponds to transition probability from hidden state ``c'' to hidden state ``i'' (that is, a $q$-gram model yield a $q-1$-th order HMM);
    \item the \emph{emission probabilities} depend on the precision and recall of the classifier on each class: for instance, if a certain symbol $s_1$ is frequently mislabelled with $s_2$, then the probability for the hidden state $s_1$ of emitting $s_2$ has to be set accordingly.
\end{itemize}

\begin{figure}
    \centering
    \subfloat[c,e]{\includegraphics[width=0.09\textwidth]{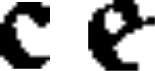}}
    \hspace{1cm}
    \subfloat[i,r]{\includegraphics[width=0.09\textwidth]{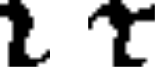}}
    \hspace{1cm}
    \subfloat[d,o]{\includegraphics[width=0.09\textwidth]{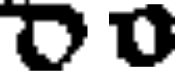}}
    \caption{Sample character pairs that can be confused due to their similar shape, or to noise in the \polygonalsegmentation{} or in the pre-processing phases, leading to punctual errors (akin to misspelling) in the generated transcriptions. \label{fig:resembling}}
\end{figure}

We also solve a decoding problem, although we observe two specific features of our setting.
\begin{itemize}
    \item We use up to $8$-grams for computing word probability, which translate into $7$-th order HMM. Unfortunately, we observed that even a $2$-nd order version of the Viterbi algorithm~\cite{viterbi} can take up to $3$ seconds for each generated transcription.
    \item Only few components of our confusion matrix have value significantly above $0$. Such components are in correspondence of character pairs with similar shapes, which we call \emph{counterparts}.
    We show in Figure~\ref{fig:resembling} sample occurrences of counterparts, that can be challenging for humans too.
\end{itemize}

For the above reasons, we trade off model accuracy for efficiency. We implement a fast procedure which optimizes the same objective function of the Viterbi algorithm, under a simplifying assumption on emission probabilities which makes sense in our setting. Specifically, we assume that the probability that the hidden symbol $h$ emits the observed symbol $o$, i.e., $p(o | h)$, is either $0$ or $\frac{1}{k}$, where $k$ is the number of counterparts of $h$ (included $h$ itself). Sets of counterparts are decided both by selecting the higher components of the confusion matrix, and by empirical experience. Our procedure takes as input a generated transcription and returns all the alternative transcriptions obtained by counterparts permutation, in non-increasing order of word probability. Complete list of counterpart classes in our experiments are: ``i'' and ``r'', ``o'' and ``d'', ``n'' and ``m'', ``l'' and ``f'', ``c'' and ``e''.

\begin{example}
Consider the wrong transcription ``dito'' in Example~\ref{ex:paths}, and let (for sake of this example) ``a'' be counterpart of ``i'', and ``c'' be counterpart of ``o''. Alternative transcriptions are ``dito'' (itself), ``dato'', ``ditc'', and ``datc''. In case ``dito'' were the only generated transcription from previous phases, the decoding step would have considered revising it into ``dato'', which is correct.
\end{example}

\paragraph{Final transcription set} We solve our specific decoding problem on the top $m$ transcriptions generated in the lattice phase, and produce further $M$ transcriptions. Among those $m+M$ transcriptions, we select $m$ new transcriptions in non-increasing order of word probability. We observed experimentally that transcriptions from the lattice are neither over- or under-represented in the final set. %

\section{Experiments}
\label{sec:experiments}
Our experiments were performed on a machine with an Intel Core i7-4980HQ CPU with 8 cores running at 2.80GHz, and 16GB RAM. Neural network training and prediction were performed on an Nvidia GeForce GTX 980M GPU, with CUDA Toolkit v9.0 and cuDNN v7.0. The operating system was Ubuntu 16.04 LTS, with kernel version 4.4.0-112-generic, and Python version 3.6.3. We implemented the neural network using TensorFlow 1.3, and the character-level language model using KenLM.

\subsection{Dataset}

\paragraph{Training data}  We use annotations from $2$ pages of Vatican Register $12$. This results in approximately $15 K$ characters. %
Characters with less than $1$K examples were augmented to match the required quantity and balance the training set. The augmentation process involves slight random rotation, zooming, shearing and shifting, both vertical and horizontal. The final dataset comprises $23$K examples evenly split between $23$ classes.

\paragraph{Test data} We test our system on four pages belonging to the same Vatican Register, but spanning different ages and writers, transcribed entirely by volunteer paleographers. %
After undergoing the pre-processing and \polygonalsegmentation{}, each word is transcribed independently by the system. Our system currently considers only the word images without \emph{abbreviated forms}. This is further discussed in Section~\ref{sec:conclusions}. %

\paragraph{Corpus for language model} Our corpus is composed of 716 ancient Roman Latin texts, spanning different ages and subjects, for a total of over 14M words. %
It is worth observing that the Latin language used in the Vatican Registers exhibits some differences with the ancient Roman Latin, which is typically used in publicly available corpora. These differences introduce some drawbacks, that we are currently overcoming, as we discuss later in Section~\ref{sec:conclusions}.

\subsection{Results}

We now compare different configurations of our transcription system. Our experiments are summarized below.
\begin{itemize}
    \item We show the benefit of the \polygonalsegmentation{} over the basic \oversegmentation{}, in the \textbf{character recognition} step.
    \item We compare different settings of the parameters used for the lattice creation and the \textbf{candidate transcriptions} steps.
    \item We show results for different $q$-gram sizes in the language model and measure how close are the top $3$ ranked transcriptions to the exact transcription.
    \item Finally, we quantify the benefit of the \textbf{word decoding} phase.
\end{itemize}

\noindent{} For the considered configurations, performance of our system are measured by the metrics below. 
\begin{itemize}
    \item We define the \emph{reciprocal rank} of a word image as the multiplicative inverse of the rank of the correct transcription: $1$ for first place, $\frac{1}{2}$ for second place, and so on. The \emph{Mean Reciprocal Rank} (\textbf{MRR}) is the average of the reciprocal ranks of word images in our test set.
    \item We define the \emph{Mean Word Processing Time} (\textbf{MWPT}) as the average time taken by our system for returning top $m$ transcriptions of word images in our test set.
    \item We define the \textbf{$m$-precision} as the fraction of word images in our test set, for which the exact transcription is i) generated by our system, and ii) ranked in the top $m$ positions. Classical definition of precision is captured by $1$-precision.
    \item For the top few transcriptions, we also provide edit distance statistics (\textbf{ED}) with respect to the exact transcription. \remark{Specifically, we use the \emph{Levenshtein distance} metric~\cite{edit}.}
\end{itemize}

\noindent{} Results of our experiments are computed in the default parameter setting and without word decoding, unless specified otherwise.

\begin{figure}
    \centering
    \includegraphics[height=3.7cm]{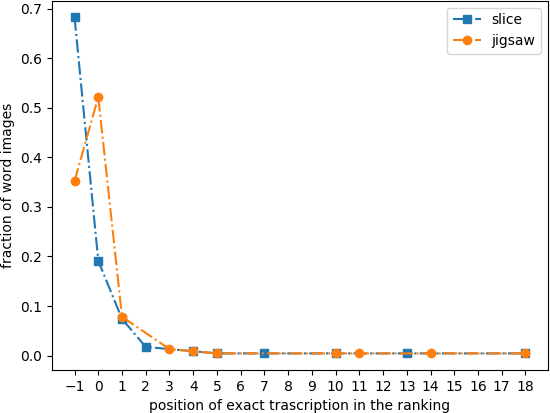}

    \caption{Comparison of different segmentation techniques. The value $-1$ represents words images for which the exact transcription has not been generated. %
    }
    \label{fig:oldvsnew}
\end{figure}

\begin{figure}
    \centering
    \includegraphics[height=1 cm]{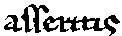}
    \caption{\remark{Exact transcription of this word image is ``asseritis'', whereas first-ranked transcription is ``afferitis''.}}
    \label{fig:editdistance}
\end{figure}

\paragraph{Segmentation} Figure~\ref{fig:oldvsnew} shows the performance of our system with different segmentation techniques. The plot considers the position in the ranking of the exact transcription of word images in our test set. The value $-1$ is a special index for when the exact transcription has not been generated. We observe that the fraction of words for which \polygonalsegmentation{} yields the exact transcription is approximately $65\%$ (decoding can recover the exact transcription of approximately $9\%$ of the remaining $35\%$), compared to much lower $20\%$ achieved by \oversegmentation{}. Also, transcriptions generated by \polygonalsegmentation{} let language model rank the exact transcription of almost all the word images\footnote{for which exact transcription is available.} in the top $5$. For remaining $35\%$, $16\%$ of first-ranked transcriptions is at edit distance $1$ from exact transcription, $15\%$ at distance $2$ and $28\%$ at distance $3$. \remark{Figure~\ref{fig:editdistance} shows a sample word image of the $15\%$ group, for which first-ranked transcription is at distance $2$ from the exact transcription.}

\begin{figure*}
    \centering
    \subfloat[MRR for $\eta$, $\theta_1$ and $\theta_2$ \label{fig:parametersA}]{\includegraphics[height=3.7cm]{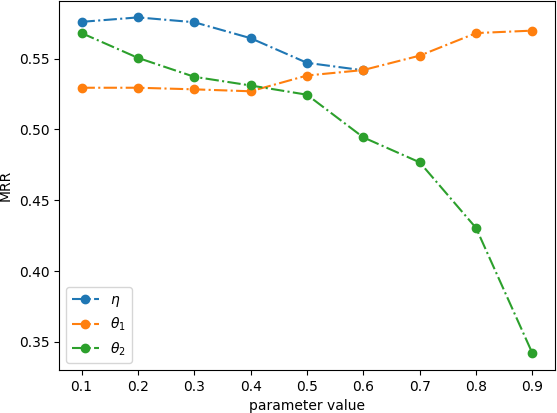}}
    \hspace{0.1cm}
    \subfloat[MWPT for $\eta$, $\theta_1$ and $\theta_2$ \label{fig:parametersB}]{\includegraphics[height=3.7cm]{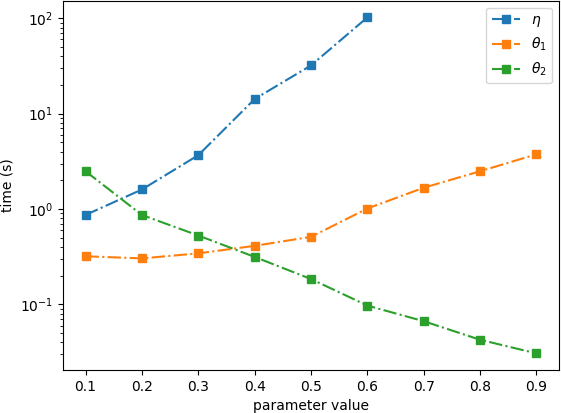}}
    \hspace{0.1cm}
    \subfloat[MRR and MWPT for $\beta$\label{fig:pruning}]{\includegraphics[height=3.7cm]{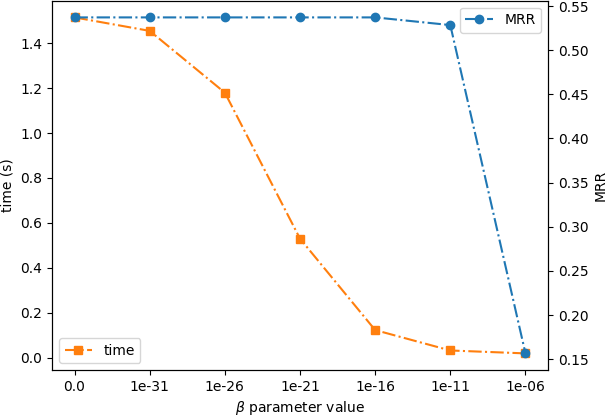}}
    \caption{Different parameter settings for the lattice creation and the transcriptions generation. Computation for $\eta>0.6$ exceeded $2$ min timeout. In Figure~\ref{fig:pruning}, we show MRR and MWPT with shared $x$-axis (in log-scale). The value $0$ is not on scale.}
\end{figure*}

\paragraph{Parameters tuning.} Figures~\ref{fig:parametersA} and~\ref{fig:parametersB} show the performance of our system for different settings of the parameter for lattice creation, that are, $\eta$, $\theta_1$ and $\theta_2$ (defined in Section~\ref{sec:recognition}). Changing such parameters affects indeed the \emph{size} -- that is, the number of edges and labels -- of the transcription lattice, trading off quality of the transcription for computing time. Dropping edges with non-character probability $> \eta$ has great impact on performances, and the smaller the threshold the better. We recall indeed that $\otimes$ labels (i.e., non-character) returned by our classifier are correct in the $95 \%$ of the tested cases\footnote{That is, false positive are very infrequent}. Dropping edges with character probability $< \theta_2$ can decrease running time akin to $\eta$. However, faster computation has a quality  cost here, and we identify the best MRR-MWPT trade-off to $\theta_2=0.1$. $\theta_1$ has little impact, due to skew in the probability distribution returned by our character classifier. After building the lattice, the parameter $\beta$ limits the number of paths that are visited during candidate transcriptions generation, as detailed in Section~\ref{sec:generation}. Figure~\ref{fig:pruning} shows that MRR stays high up to $\beta=10^{-11}$ and drops afterwards. $\beta=10^{-16}$ has best MRR-MWPT trade-off.

\begin{figure}
    \centering
    \subfloat[$1$-precision\label{fig:prec1}]{\includegraphics[height=3.7cm]{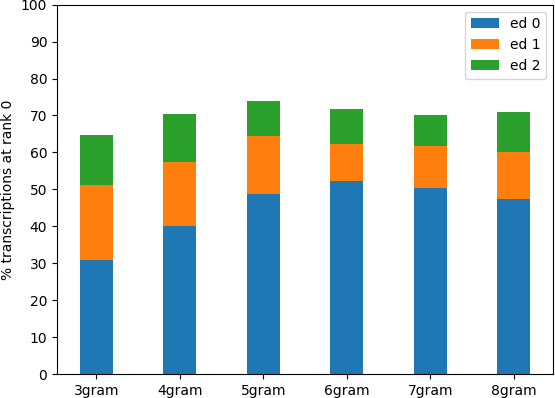}}

    \subfloat[$3$-precision\label{fig:prec3}]{\includegraphics[height=3.7cm]{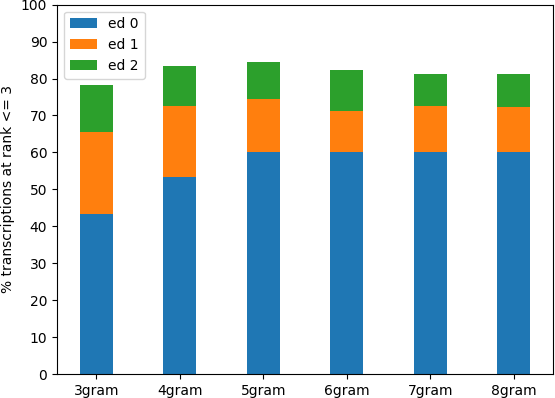}}
    \caption{Different values of $q$ in the language model.}
    \label{fig:ngramcomparison}
\end{figure}

\paragraph{Ranking transcriptions} The blue bars in Figure~\ref{fig:ngramcomparison} show the $1$-precision and $3$-precision of our system for different $q$-gram sizes in the language model. Figure~\ref{fig:prec3} considers top $3$ transcriptions of all the word images in the test set. We observe that, by using $6$-grams, almost $80\%$ of our results are away from exact transcriptions by no more than $2$ characters, and approximately $60\%$ corresponds exactly to the underlying manuscript word.
Figure~\ref{fig:prec1} reports the corresponding results when considering only top $1$ transcriptions. Another way for reading results in Figure~\ref{fig:prec1} is the following. As shown in previous Figure~\ref{fig:oldvsnew}, we generate the exact transcription of more than $65\%$ of the word images in our dataset. This means that approximately $77\%$ of such word images have the exact transcription ranked at position $1$ when using $6$-grams (which is our default setting), but approximately $23\%$ does not get the optimal ranking\footnote{Exact transcription, when generated, is in the top $5$ for almost all the word images.}. Improving on the ranking produced requires a better model of the language used in the Vatican Register, included models of sentences, and is discussed in Section~\ref{sec:conclusions}.%

\paragraph{Fast decoding} We observe that the fraction of word images for which our system does not generate the exact transcription is approximately $35\%$. (It is worth noticing that for such word images, most of first-ranked transcriptions have no more than $3$ spelling errors.) Decoding can recover the exact transcription of approximately $9\%$ of such word images. Other effects of the decoding phase is that top-ranked transcriptions become closer to exact transcriptions. For instance, the amount of word images having exact transcription ranked as second increases by $30\%$. We note that the amount of word images having exact transcription ranked as first does not change significantly. Overall, only top-ranked positions benefit from decoding. Indeed, decoding has limited effect on the overall MRR, which changes from $0.57$ to $0.58$ by using decoding.

\section{Related Works}
\label{sec:related}

Handwritten Text Recognition (or HTR) is a branch of research concerning the automatic transcription of handwritten text. 
Even though this it extends to live-captured handwriting (online recognition), that is clearly not the case for historical documents. The offline recognition task is generally regarded as harder than the online, due to the lack of temporal information: online handwriting recognition can leverage the order and timing of character strokes, while this is not the case for offline recognition. 
As more and more libraries and archives worldwide digitize their collections, great effort is being put into the creation of full-fledged off-line HTR systems~\cite{sanchez2014icfhr2014,flaounas2013research,sanchez2014handwritten}. 
These transcription systems generally work by a segmentation-free approach, where it is not necessary to individually segment each character. While this removes one of the hardest steps in the process, it is necessary to have full-text transcriptions for the training corpus, in turn requiring expensive labelling procedures undertaken by paleographers with expertise on the period under consideration. \cite{fischer2010ground} details all the steps for the creation of the IAM-HistDB, one of the first publicly available historical handwritten text databases containing medieval script.
Our character-level classification has instead much smaller training cost, and allows the collection of a large corpus of annotated data using a cheap crowdsourcing procedure.

Due to the many challenges involved in a fully automatic transcription system of historical handwritten documents, many researchers in the last years have focused on solving sub-problems, including word spotting \cite{rath2007word,puigcerver2015icdar2015},
handwriting classification \cite{bulacu2007automatic}, and text line segmentation \cite{likforman2007text}.

Crowdsourcing solution for cultural heritage has been experienced in many projects. One of the pioneering initiative to crowdsource the transcription of manuscripts is the \emph{Transcribe Bentham} project, a collaborative platform for crowdsourcing the transcription of the philosopher Jeremy Bentham's unpublished manuscripts~\cite{causer2014many}. Also the Transcriptorium project~\cite{sanchez2013transcriptorium,marcus2015crowdsourced} exposes HTR tools through specialized crowd-sourcing web portals, supporting collaborative work. 
Our solution is more focused than the above ones: since it aims at producing training data, it relies on a much simpler solution based on visual pattern matching task that can be performed by unskilled workers.

There has been an interest in applying recent results in recurrent and convolutional neural networks to achieve improved classification accuracy: \cite{sudholt2016phocnet} performs word spotting through a deep convolutional neural network, outperforming various word spotting benchmarks; while \cite{fischer2009automatic} adopts a bidirectional Long Short-Term Memory neural network to transcribe at word level, with high accuracy. We will come back on this point when discussing future research directions for our project.

A variety of preprocessing steps as those applied in our system are discussed in~\cite{kim2014online,Jaeger:2003:SAJ:2722884.2722986}. Common steps include normalization of size, density, rotation (slope), and slant. HMM technology is widely used in segmentation-free recognizers~\cite{koerich2003large,toselli2004spontaneous,Romero:2007:CNS:1417835.1417895,Bertolami20083452}. Language modeling for on-line handwriting recognition bears many similarities with the use of language information in OCR and speech recognition, which often employ statistical $n$-gram models on the character or word level~\cite{Bazzi:1999:OOO:305482.305484,stolcke2002srilm}.

\section{Conclusions and future work}
\label{sec:conclusions}

\begin{figure}
\includegraphics[height=1cm]{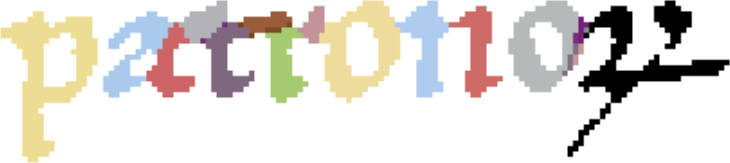}
\caption{A word (\emph{patrono}rum) from the Vatican Registers containing an abbreviation (the symbol in black).\label{fig:patronorum}}
\end{figure}

Data science can deeply contribute to analyze and understand our historical and cultural heritage. Data acquisition and preparation from manuscript historical documents is done by means of a transcription process, whose scalability is limited, as it must be performed by expert paleographers. In this paper, we have presented a system, developed in the context of the \incodiceratio{} project, to support the transcription of medieval manuscripts in order to improve the scalability of the process. We have followed an original approach that requires minimal training efforts, and that is able to produce correct transcriptions for large portions of the manuscripts. Our approach, which relies on word segmentation, neural convolutional network, and language models, has been successfully experimented on the Vatican Registers.

\smallskip
We are currently working on the system in order to extend the set of symbols, and hence to improve the overall effectiveness of the process. In particular, we are adding the most frequent abbreviations, i.e., short-hands used by the scribes to save room or to speed up writing. 
In our process, the main issue with abbreviations is the lack of statistics on their occurrences, which prevents us to effectively apply the language models. Gathering statistics for the abbreviation is not trivial: the usage of these symbols depends both on the age and on the domain of the manuscripts. For instance, in the Vatican Registers, which have diplomatic and legal contents, some abbreviations are more frequent than in manuscripts with of literary works, even from the same age.  Indeed, we have already collected training samples for the classifier also for many abbreviations: our crowdsourcing approach to collect labeled examples worked well also for these symbols, as it is based on simple visual pattern matching tasks. Figure~\ref{fig:patronorum} shows an example of a one of the most frequent abbreviations. The last symbol, in black, is a shorthand for the Latin desinence \emph{``rum''}: notice that it is simple, given some sample images, to recognize it also without any paleography knowldge. Also the neural network performs well with the extended set of symbols, as abbreviations are typically well distinguishable from other symbols. 

Our plan to collect statistics for the abbreviations is to use our current system to produce partial transcriptions for a number of pages, a few dozens, highlighting the words where the character classifier recognizes an abbreviation. Then, we will ask to the paleographers to transcribe these words. Based on these semi-automatic transcriptions, we will progressively update the language models.

Another direction that we are currently exploring is to change the human interaction interface of the crowdsourcing platform. The idea is to show a complete word, segmented by the \polygonalsegmentation{} method, and to ask to the worker to highlight the segments corresponding to a target symbol. We expect to obtain more accurate examples, which could improve the accuracy of the neural network, especially for the class of non-character $\otimes$. Also, these annotations could allow us to build synthetic manuscripts, which could be used to train a recursive neural network.

\section*{Acknowledgments} We thank Andrea Rossi for his fundamental contribution on a earlier version of the system and Simone Scardapane for his work on the deep convolutional neural network. We thank Gaetano Bonofiglio, Veronica Iovinella e Andrea Salvoni for their work on a earlier version of the classifier. We thank Matteo Mariani for helping with all the pre-processing steps. We thank Debora Benedetto, Elena Bernardi and Riccardo Cecere for collaboration on \polygonalsegmentation{} and decoding. We also thank Massimo Bove and Gianlorenzo Didonato for developing the front-end of the \incodiceratio{} crowd-sourcing application. We thank prof. Serena Ammirati and the paleoghapers Marco Miglionico and Michela Ventriglia, who volunteered for transcribing our test set. We are indebted to prof. Marica Ascione and to the students of Liceo Keplero and Liceo Montale of Rome who did the labelling effort. We thank S\'ebastien Brati\`eres and Lukasz Kaiser for helpful discussion, and the staff of Pi School\footnote{\url{http://picampus-school.com}} for supporting the author Elena Nieddu. We gratefully acknowledge the support of NVIDIA Corporation with the donation of a Quadro M5000 GPU for future steps of this research.

\bibliographystyle{abbrvurl}
\bibliography{references}

\end{document}